\documentclass[aps,prl,twocolumn,groupedaddress,reprint]{revtex4-1}

\usepackage{graphicx}
\usepackage{natbib}

\begin{document}


\title{Electron phonon coupling versus photoelectron energy loss at the origin of replica bands in photoemission of FeSe on SrTiO$_3$}

\author{Fengmiao Li and George A. Sawatzky}

\affiliation{Department of Physics \& Astronomy, University of British Columbia, Vancouver, British Columbia, Canada V6T 1Z1}
\affiliation{Stewart Blusson Quantum Matter Institute, University of British Columbia, Vancouver, British Columbia, Canada V6T 1Z4}

\date{\today}

\begin{abstract}
The recent observation of replica bands in single-layer FeSe/SrTiO$_3$ by angle-resolved photoemission spectroscopy (ARPES) has triggered intense discussions concerning the potential influence of the FeSe electrons coupling with substrate phonons on the superconducting transition temperature. Here we provide strong evidence that the replica bands observed in the single-layer FeSe/SrTiO$_3$ system and several other cases are largely due to the energy loss processes of the escaping photoelectron, resulted from the well-known strong coupling of external propagating electrons to Fuchs-Kliewer (F-K) surface phonons in ionic materials in general. The photoelectron energy loss in ARPES on single-layer FeSe/SrTiO$_3$ is calculated using the demonstrated successful semi-classical dielectric theory in describing low energy electron energy loss spectroscopy of ionic insulators. Our result shows that the observed replica bands are mostly a result of extrinsic photoelectron energy loss and not a result of the electron phonon interaction of the Fe d electrons with the substrate phonons. The strong enhancement of the superconducting transition temperature in these monolayers remains an open question.
\end{abstract}


\maketitle

The discovery of enhanced superconductivity with transition temperatures (T$_C$) up to $\sim$65~K in the single-layer FeSe/SrTiO$_3$ \cite{qiwang-cpl,he2013phase} has drawn intense discussion in the condensed-matter community regarding the origin of the T$_C$ enhancement. The cross-interface electron-phonon coupling is widely regarded to play an important role in this \cite{wang2016interface,huang2017monolayer} and the acclaimed strongest evidence is the observation of ``shake-off'' replica bands by angle-resolved photoemission spectroscopy (ARPES) at an energy of 90-100~meV \cite{lee2013interfacial} in the range of a strong optical phonon in the SrTiO$_3$ (STO) substrate. It is indeed tempting to attribute the replica bands to the coupling of FeSe electrons with STO optical phonons \cite{wang2016interface,huang2017monolayer,lee2013interfacial,lee2015makes}; however, theoretical estimates of the strength of this coupling and contribution to T$_C$ find this to be at most a small contribution \cite{rademaker2016enhanced,wang2016aspects,ywang2016,yzhou2017,gl2016}. In subsequent experiments, similar replica bands and enhanced superconductivity properties were observed on single-layer FeSe films grown on different substrates such as BaTiO$_3$(001) \cite{peng2014tuning}, STO(110) \cite{zhang2017ubiquitous} and rutile TiO$_2$(100) \cite{rebec2017}. This seems to suggest a commonality and an intrinsic nature of the replica bands. However, to get narrow replica bands tracing closely the dispersion of main bands crossing the Fermi energy rather than broad ``shake-off'' features or kinks, one has to require the coupling to be strongly peaked at $q_{\parallel}=0$.  Arguments for this strong peakedness were given in references \cite{lee2013interfacial,rademaker2016enhanced,wang2016aspects}. We will show that the surface Fuchs-Kliewer (F-K) phonon \cite{fuchs1965,ibach1970,ibach1982electron} coupling to the escaping photoelectron is naturally strongly peaked at $q_{\parallel}=0$ and in comparison to electron energy loss spectroscopy we provide direct evidence that the replica bands are a consequence of energy loss processes of the escaping photoelectron. Our conclusions are of general importance in the interpretation of self energies in ARPES measurements in insulators or low carrier density metals.

Enlightening in the search for an explanation of the replica bands in ARPES is that replica bands at the same ``shake-off'' energy are observed also in  bare STO surfaces exhibiting a two-dimensional electron gas (2DEG) in ARPES \cite{zhang2017ubiquitous,chen2015observation,wang2015tailoring}. Even more enlightening is the recent report by Zhang \textit{et al} \cite{FeSe-eels} of strong energy loss features in high-resolution electron energy loss spectroscopy (HREELS) at the same energies as the replica bands on a single-layer FeSe/STO(001) surface as well as on the bare STO surface. The HREELS results as we will show can be accurately described by semi-classical dielectric theory without adjustable parameters. This provides strong evidence that these loss features are due to the strong coupling of the electron approaching and then reflecting from the surface, with the deeply penetrating F-K surface phonon modes \cite{fuchs1965,ibach1982electron}. The F-K phonons with $q_{\parallel}=0$ involve modes of the positive and negative ions opposite motion perpendicular to the surface extending deep into the bulk of ionic material \cite{fuchs1965,ibach1970,ibach1982electron,thiry1984,liehr1984high,kesmodel1981,baden1981,conard1993}. This produces a large oscillating potential extending far outside the surface. These couple strongly with electrons moving with a component perpendicular to the surface resulting in the energy loss but little parallel momentum loss \cite{evans-1972}.

ARPES is one of the most direct and widely-used methods to study the energy and momentum dispersion relation and especially also the so-called real and imaginary parts of the self-energy resulting from the interaction of the electrons with each other and with bosonic degrees of freedom involving excitations like magnons, excitons and phonons \cite{damascelli-arpes}. Photoemission, however, is a two-particle excitation involving the photoinduced hole and the photoelectron [Fig. 1(a)] and even if the electron-hole interaction is small the two particles each carry a self-energy due to the interaction with the rest of the system. Mostly we assume, and rightly so, that if the energy of the photoelectron is very high, \textit{i.e.}, high photon energy, the interaction with the photoinduced hole can be neglected (the so-called ``sudden approximation''). We also mostly assume that the interaction of the escaping photoelectron with the sample can be neglected. In this case the spectra obtained will provide direct information concerning the self-energy of the photoexcited hole. However, in the case of ionic materials with F-K phonons the huge oscillating long-range external potentials above the surface \cite{ibach1982electron,lucas1972fast} can interact strongly with the photoelectron after it escapes from the sample and travels to the analyzer.

The strong interaction of externally moving electrons have been very clearly demonstrated by the HREELS experiments on ionic material surfaces for example on ZnO surface by Ibach \cite{ibach1970}. The energy loss peak height sometimes can be even 50\% of the zero-loss peak height when using incident energies in the same range as often occurs in the kinetic energy of emitted electrons in ARPES. In this case, the collected ARPES spectrum must be corrected for the energy loss processes of the emitted photoelectron. However, to our knowledge, this information from HREELS has not been considered and used to correct ARPES spectra generally.

In this Letter, we study the photoelectron energy loss process on the single-layer FeSe/STO(001) system due to photoelectrons interacting with the F-K phonons of STO substrate. We use the well-developed semi-classical dielectric theory to describe HREELS on the STO surface which depends only on the optical constants of the material and the electron kinetic energy and propagation direction, and also describe the energy loss processes of the emitted photoelectron electron in ARPES. The optical constants are available from infrared optical spectroscopy studies of STO \cite{servoin-1980,galzerani1982infrared}.

\begin{figure}[b]
\includegraphics[clip,width=2.7 in]{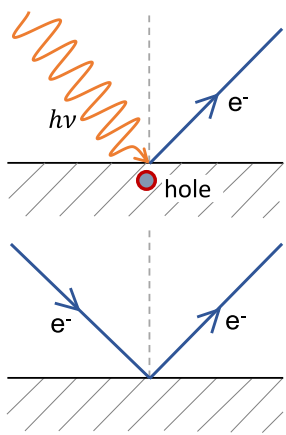}
\caption{A comparison of the angular resolved photoemission (top) and high resolution specular reflection electron energy loss spectroscopy processes (bottom).}
\end{figure}

ARPES involves a photon-in-electron-out process [Fig.~1, top] and most experiments are conducted with less than 100~eV photon energy or photoelectron kinetic energy about the work function lower than the photon energy when looking at states close to the chemical potential in metals \cite{damascelli-arpes}. For low energy losses as for phonons of less than 100~meV, the trajectory of the emitted photoelectron hardly changes. HREELS involves an electron-in-electron-out process [Fig.~1, bottom] with the monochromatized incident electron energy usually less than 100~eV \cite{ibach1982electron}. Since in specular reflection the total path of the electron in HREELS is twice as long as in ARPES and realizing that interference terms between the loss processes for the incident and reflected electrons are small except for grazing incidence \cite{lucas1972fast}, we can safely approximate the intensity of the loss in ARPES to be $\frac{1}{2}$ of that in HREELS for the same kinetic energy and same trajectories. This is the main concept we use in what follows.

The F-K surface optical phonon modes was first predicted by Fuchs and Kliewer \cite{fuchs1965} to be present in ionic crystal slabs and somewhat later was detected by H. Ibach on ZnO in a  HREELS study \cite{ibach1970}. In HREELS, the impinging electron generating a time variable electric field normal to the surface direction can interact strongly with the long-wavelength surface optical phonon  \cite{ibach1982electron,lucas1972fast}. As shown in Fig. 2(a), the vibrating cation and anion with $q_{\parallel}=0$ generate an oscillating potential in the vacuum which for an infinitely extending surface is independent of the distance from the surface like the static case of a polar terminated ionic crystal (\textit{i.e.}, the ``polar catastrophe''). The magnitude of the oscillating potential outside the surface is proportional to the penetration depth of the ionic displacements into the bulk of the crystal which decays exponentially with $q_{\parallel}$ \cite{ibach1982electron,evans-1972,evans-1973}. Therefore the $q_{\parallel}=0$ mode is by far the strongest. The huge long-range dynamic potential results in the very large inelastic-scattering probability in the close to specular reflection direction in HREELS \cite{ibach1970,ibach1982electron,thiry1984,liehr1984high,kesmodel1981,baden1981,conard1993} . Since the incident electron energy is much larger than the optical phonon energy, \textit{i.e.}, ``fast'' incoming electrons, classical electromagnetic theory describes the energy loss process accurately resulting in the single-loss probability at zero temperature given by \cite{ibach1982electron,ibach1994116}
\begin{equation}
P(w)=\frac{2}{a_0k_i\cos\theta_i}\frac{1}{w}Im(\frac{-1}{\epsilon_b(w)+1})
\end{equation}
where $a_0$ is the Bohr radius, $k_i$ and $\theta_i$ are the wavevector and incident angle of the incoming electron relative to the perpendicular surface direction, respectively, and $\epsilon_b(w)$ is the frequency dependent bulk dielectric function. For an isotropic material with $n$ infrared-active transverse optical phonons, the dielectric function in the Lorentz model is given by \cite{ibach1982electron}
\begin{equation}
\epsilon_b(w)=\epsilon(\infty)+\sum^n_{k=1}\frac{Q_kw^2_{TO,k}}{w^2_{TO,k}-w^2-i\gamma_kw}
\end{equation}
where $\epsilon(\infty)$ is the high-frequency dielectric constant, $w_{TO,k}$ the infrared-active transverse-optical phonon frequency, $Q_k$ the oscillator strength and $\gamma_k$ the damping frequency. These crystal properties can be obtained directly from the infrared optical experiments. In the case of only one oscillator, $Re(\epsilon)=-1$ in Eq.~(2) determines the F-K phonon energy at $w_{FK}=w_{TO}\sqrt{\frac{\epsilon(0)+1}{\epsilon(\infty)+1}}$ where $\epsilon(0)$ is the static dielectric constant \cite{ibach1982electron}. The F-K phonon energy lies between the transverse and longitudinal optical phonon frequencies using the Lyddane-Sachs-Teller relation $\frac{w_{LO}}{w_{TO}}=\sqrt{\frac{\epsilon(0)}{\epsilon(\infty)}}$. For large inelastic-scattering probabilities, perturbation theory breaks down and the quantum-mechanical harmonic oscillator introduced by Lucas and Sunjic \cite{lucas1972fast,lucasprl1971,lambin1990computation} describes the amplitude of the multiple scatterings exhibiting a Poisson intensity distribution in the ideal case of a sole, undamped excitation at zero temperature.

The semi-classical dielectric theory was justified by Evans and Mills \cite{evans-1972,evans-1973} using the completely quantum mechanical method and HREELS spectra simulated with the semi-classical dielectric theory including the Poisson distribution are in very good agreement with experiments for many ionic material surfaces \cite{thiry1984,liehr1984high,kesmodel1981,lucasprl1971,matz1981} and in the following we will use this theory to simulate the energy loss process in the HREELS and ARPES experiment. Figure~2(b) shows very intense energy-loss peaks in the HREELS measured on the STO(001) surface at 470~K by Conard \textit{et al}. \cite{conard1993}. The two strongest energy-loss peaks are at $\sim$59~meV (w$_{FK1}$) and $\sim$92~meV (w$_{FK2}$) with multiple-phonon scatterings at $\sim$118~meV (2w$_{FK1}$), $\sim$149~meV (w$_{FK1}$+w$_{FK2}$) and $\sim$184~meV (2w$_{FK2}$). The energy gain peaks on the left of zero energy loss is due to the interactions with the thermal excited phonon \textit{i.e.}, the ``anti-Stokes peak''. The STO dielectric function given by Eq.~(2) can be approximated using  $\epsilon(\infty)=5.5$ and the oscillator parameters directly obtained from the infrared optical experiment \cite{supplemental}--at 470~K, three infrared-active transverse optical phonons including the soft mode TO1 at $\sim$14~meV (Ti-O-Ti bending), TO2 at $\sim$22~meV  (Sr ion moving against the TiO$_6$ octahedra) and TO4 at $\sim$67~meV (the Ti-O stretching) \cite{servoin-1980,perry1964,axe1967}. At low temperature, a weak infrared-active mode due to the rotation of the neighboring oxygen octahedral is included \cite{galzerani1982infrared}. As shown in Fig.~2(b), the simulated HREELS spectrum using the semi-classical dielectric theory exhibits very good agreement with experiment in both energy gain and loss side including also the relative intensities. We note that the inelastic scattering on STO surface is dominated by the soft mode \cite{conard1993}. At low temperature, it is well known that the TO1 mode in STO becomes soft dramatically with a very large STO static dielectric constant but interestingly the calculated [Fig.~2(b)] and experimental HREELS at low temperature \cite{FeSe-eels} changes little from the 470~K data aside from the gain peaks. The reason for this small change despite the large change in the static dielectric constant is that with variable temperatures the soft mode keeps an almost constant $Qw_{TO}^{2}$\cite{Barker-1966,Tinkham-1962} which determines the F-K phonon energy and intensity \cite{supplemental}.

\begin{figure}[t]
\includegraphics[clip,width=3.0 in]{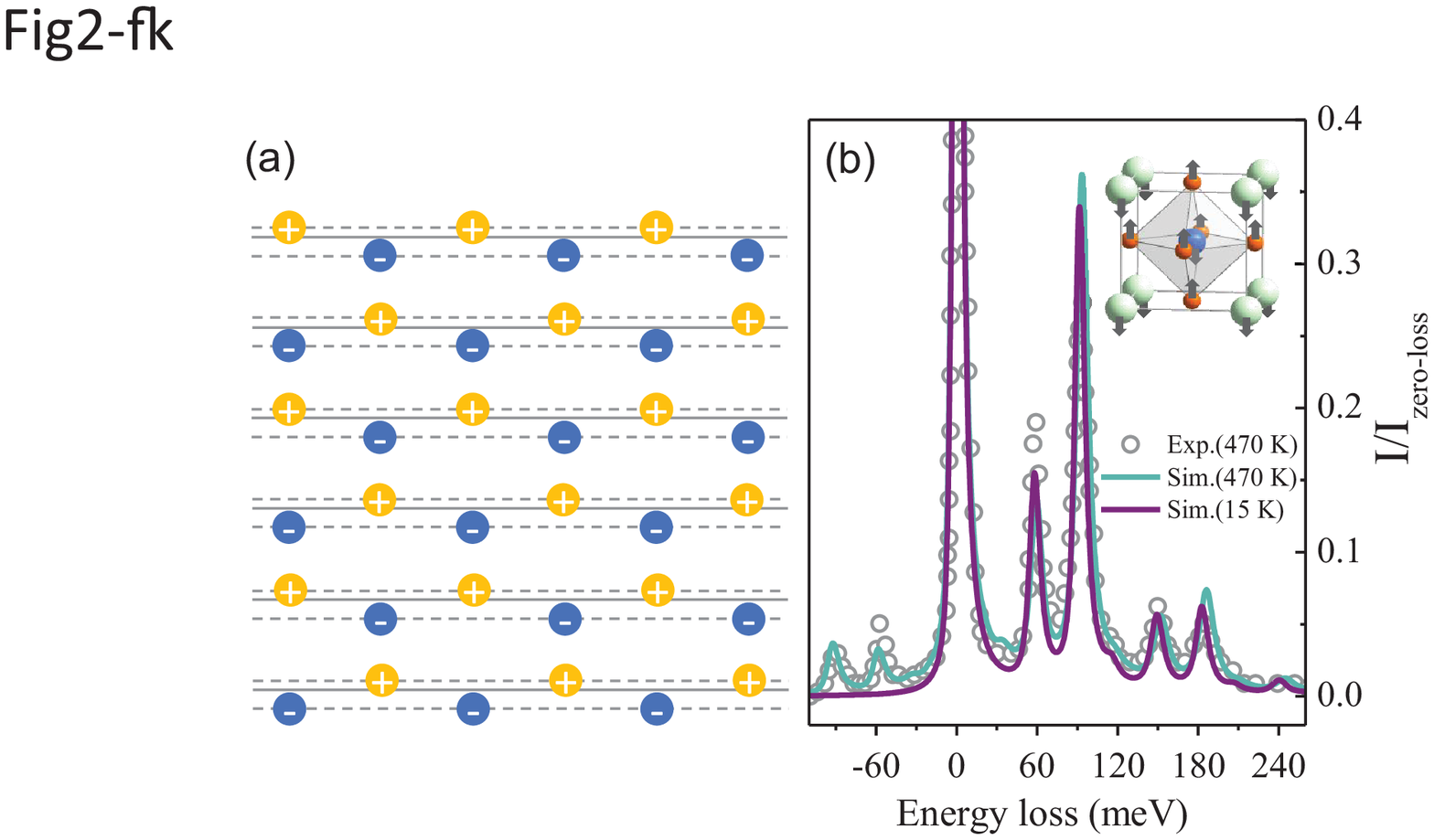}
\caption{(a) A snapshot of the instantaneous position of cation and anion displacements for the long-wavelength surface optical phonon in a cut out side view of a (001) terminated rocksalt structure slab. The solid lines show the equilibrium positions of the ions. Here the anions are assumed to be much lighter than the cations. (b) The experimental (at 470~K) \cite{conard1993} and simulated HREELS (at 470~K and 15~K) of STO(001) terminated crystal with 6.0~eV incoming electron and incident angle $\theta=45^o$.The inset illustrates the polar displacement in STO.}
\end{figure}

In the following, photoelectron energy losses in ARPES on the single-layer FeSe/STO(001) system will be studied. We consider the high resolution very clear experimental data of the Stanford group on a well characterized single-layer FeSe film grown on a 0.05~wt\% Nb-doped STO substrate \cite{lee2013interfacial}. The Nb doping that is needed to prevent any charging effects in ARPES also results in a low-density electron gas in the STO which will result in some screening of the F-K phonon modes which will tend to decrease the experimental replica features relative to theory done for the insulating case. Free electrons in the substrate can couple with the F-K phonon and result in a shift and lower intensity \cite{ibach1982electron,matz1981,supplemental}. However, the effect of charge carriers in the Nb doped STO substrate and FeSe film is expected to be very small given the low Nb doping concentration and the two-dimensionality of the FeSe based electron gas. For a further discussion of the dependence on the electron density we refer to Section~3~\& 4 in the supplementary material \cite{supplemental}. The ARPES experiment is performed at 20~K using 24~eV photons \cite{lee2013interfacial} and a kinetic energy of the photoelectron around $\sim$19.5~eV. Replica bands around the M point of the 2 Fe per unit cell Brillouin zone are measured with the analyzer angle $\theta\approx31^o$ relative to the perpendicular to surface.

FeSe monolayers on STO(001) exhibit two bands labeled A and B and this results in two sets of replica structures. In Fig.~3(a), we display the calculated ARPES spectra using the optical parameters of STO at 15~K \cite{galzerani1982infrared} and using the same theory as that used for the HREELS spectra but reducing the loss features by a factor of 2 and modified for the somewhat different kinetic energies and angles according to energy loss theory described above. Strong ``shake-off'' peaks at $\sim$90~meV in Fig.~3(a) are observed in both the A and B bands due to the STO F-K phonon at $\sim$90~meV. The energy loss simulation for the A band has, in addition, a small peak with ``shake-off'' energy $\sim$60~meV corresponding to the STO $\sim$60~meV F-K phonon. As shown in the bottom of Fig.~3(a), the summation of A and B results in very good agreement with the energy distribution curve from experiment. The energy loss parts of the A and B bands [Fig.~3(b)] reproduce all the replica features observed experimentally in the reproduced data of reference \cite{lee2013interfacial} in Fig.~3(c). The C band observed in experiment is also generated in the simulation and is due to the $\sim$60~meV energy loss peak from the A band as mentioned above. The reproduction of the experiment with the semi-classical dielectric theory in HREELS and all parameters coming from optical experiments without any fitting parameter, provides very strong evidence that the replica band can be well described by photoelectron energy loss processes due to the excitation of F-K phonon in the STO substrate. This can also explain the replica on the FeSe/BaTiO$_3$ \cite{peng2014tuning} and FeSe/TiO$_2$ \cite{rebec2017} system, on which the strong electron energy loss peak at $\sim$90 meV is observed in experimental HREELS and our simulation \cite{supplemental,kesmodel1981,shuyuan-2018,cox1986high}.

\begin{figure}
\includegraphics[clip,width=3.3 in]{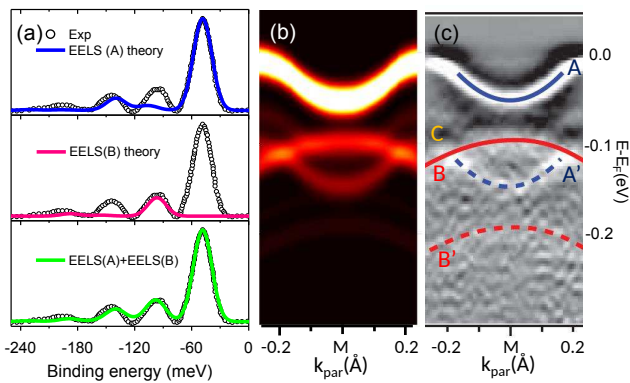}
\caption{(a) The theoretical  photoelectron energy loss resulting from band A [top, EELS(A)] and B [middle, EELS(B)] at the M point compared with the experimental ARPES result \cite{lee2013interfacial,supplemental}. The simulated spectra were shifted to match the primary A and B bands in experiment. Bottom is the comparison with the experimental curve. (b) The simulated total momentum dependent ARPES spectra including the principle and two replica bands \cite{supplemental}. The intensities calculated for the M point were assumed to be wavevector independent because of the small spread in the $\pm$0.2~\AA$^{-1}$ momentum range covered. (c) The experimental ARPES result around the M point \cite{lee2013interfacial}.}
\end{figure}

The probability of generating excitations in the energy loss process depends on measurement geometry and photoelectron velocity. As shown in Eq.~(1), it is inversely proportional to the electron momentum perpendicular to the surface, \textit{i.e.}, $k\cos(\theta)$. Although this relation is not valid for close to grazing incidence because of the strong influence from the image charge potential and the interference of the incoming and outgoing electron in HREELS, and also invalid for slow electrons which are affected by the recoil effect \cite{lucas1972fast}, the theory can be further tested with a photon energy and Brillouin zone dependent study staying within the limits of photoelectron energy $>$~10~eV and emission angle $<$~60$^{o}$ \cite{supplemental}.

The energy loss processes of escaping electrons may also be the explanation for a number of replica bands or ``peak-dip-hump'' structures observed in thin metallic capping layer on an ionic substrate such as STO \cite{lee2013interfacial,zhang2017ubiquitous,chen2015observation,wang2015tailoring}, TiO$_2$ \cite{rebec2017,moser2013} and ZnO \cite{Yukawa2016}. When the free electron density is low or highly-confined at the surface, the screening effect of the free electrons is weak and in ARPES replica bands are clearly visible \cite{supplemental}. When the free electron density is high and spatially distributed deeply, it exhibits a very low and broad peak due to strong screening by free electrons \cite{ibach1982electron,matz1981,supplemental}. The replica intensity is indeed observed in recent experiments to have a strong dependence on the free electron density on the STO surface \cite{zhang2017ubiquitous,wang2015tailoring}.

In conclusion, we have provided strong evidence that the replica bands recently observed for the single-layer FeSe/STO system are largely due to the extrinsic photoelectron energy losses by exciting STO F-K surface phonons. This explains the observed replica structure in detail without the need of any additional electron-phonon coupling due to the FeSe electrons interacting with substrate phonons. This does not necessarily eliminate the importance of electron phonon coupling of the d electrons in FeSe to the substrate phonons which, in principle, if strong could lead to small polaron formation; but, to do this one would need a very large q range in the coupling as in the Holstein model and this would result in kinks or very broad continuumlike shake up structure in ARPES.  In addition, our studies strongly suggest that corrections must be introduced when analyzing photoemission spectroscopy on ionic material surfaces. Complementing the ARPES with HREELS studies on the same systems can provide the information needed to correct for this.

\textit{Note added:} In a recent paper \cite{song2017phonon}, the oxygen isotope effect on the replica energy in ARPES and HREELS has been reported. Our energy loss calculation actually can reproduce the isotope effect \cite{supplemental}. Based on our analysis, the replica intensity does not have a direct relationship with superconductivity in this system. If the linear dependence of the superconducting gap on replica intensity is confirmed, other potential interactions or changes in structure at the interface as a function of annealing conditions which influence both the superconducting gap and the F-K phonon loss intensity would, in our view, have to be looked for in detail. 

\begin{acknowledgments}
This work was supported by Natural Sciences and Engineering Research Council of Canada, CIfAR, and the Max Planck-UBC-UTokyo  Centre for Quantum Materials.
\end{acknowledgments}

\bibliography{ref}

\end{document}